\documentclass[useAMS,usenatbib]{mn2e}

\usepackage[dvips]{graphicx}
\usepackage[dvips]{color}

\title[Type I migration in optically thick accretion discs]{Type I
migration in optically thick accretion discs}
\author[K. Yamada and S. Inaba]{K.
Yamada$^{1}$\thanks{E-mail:k-yamada@aoni.waseda.jp; Present address:Waseda University, Nishiwaseda, Shinjuku-ku, Tokyo 169-8050, Japan} and S.
Inaba$^{2}$\thanks{E-mail:satoshi.inaba@waseda.jp}\\
$^{1}$Center for Planetary Science, Kobe University, Hyogo, Japan\\
$^{2}$School of International Liberal Studies, Waseda University, Tokyo, Japan}
\begin{document}

\date{Accepted ; Received ; in original form }

\pagerange{\pageref{firstpage}--\pageref{lastpage}} \pubyear{2012}

\maketitle

\label{firstpage}

\begin{abstract}
We study the torque acting on a planet embedded in an optically thick
accretion disc, using global two-dimensional
hydrodynamic simulations. The temperature of an optically thick
accretion disc is
determined by the energy balance between the viscous heating and the
radiative cooling. The radiative
cooling rate depends on the opacity of the disc. The opacity is
expressed as a function of the
temperature. We find the disc is divided into three regions that have
different temperature distributions.
The slope of the entropy distribution becomes steep in the inner
region of the disc with the high
temperature and the outer region of the disc with the low temperature,
while it becomes shallow in the
middle region with the intermediate temperature. Planets in the inner
and outer regions move outward
owing to the large positive corotation torque exerted on the planet by
an adiabatic disc, on the other
hand, a planet in the middle region moves inward toward the central
star. Planets are expected to
accumulate at the boundary between the inner and middle regions of the
adiabatic disc. The positive
corotation torque decreases with an increase in the viscosity of the
disc. We find that the positive
corotation torque acting on the planet in the inner region becomes too
small to cancel the negative
Lindblad torque when we include the large viscosity, which destroys
the enhancement of the density in
the horseshoe orbit of the planet. This leads to the inward migration
of the planet in the inner region of
the disc. A planet with 5~Earth masses in the inner region can move
outward in a disc with the surface density of 100~$\rm{g/cm^2}$ at 1~AU when the accretion rate of a disc is smaller
than $2 \times 10^{-8} M_{\rm{\odot}}$/yr.
\end{abstract}


\section{Introduction}
Planets are thought to be formed in a protoplanetary disc around a
young star. A planet growing by the
accretion of planetesimals exchanges the angular momentum with disc
gas and moves in the disc
\citep{b5, b6}. This is called the Type~I migration of a planet. The
Type~I migration is caused by the torques acting on a planet by a
disc. The torque is composed of the Lindblad torque and the corotation
torque. The Lindblad torque is due to the two spiral density waves
excited by the planet in the disc \citep{b27, b30}, while the
corotation torque is exerted by the gas in the horseshoe region of the
planet \citep{b29, b2, b21, b22}. The sum of the Lindblad torque and
the corotation torque determines the total torque acting on the
planet.

The Lindblad torque is exerted by the disc gas at the Lindblad
resonances, which are located inside
and outside the orbit of the planet. The angular
momentum of the planet is increased
and decreased by the inner and outer Lindblad torques, respectively.
The magnitude of the negative
outer Lindblad torque is a little larger than that of the positive
inner Lindblad torque in a disc \citep{b27, b30}. The Lindblad torque
becomes negative, leading to the inward migration of a
planet.

The corotation torque exerted by the disc gas in the horseshoe orbit
of a planet depends on the
vortensity distribution and the entropy distribution of the disc gas.
The vortensity related corotation
torque is small compared with the Lindblad torque \citep{b48, b45}, on
the other hand, the entropy
related corotation torque can become larger than the magnitude of the
negative Lindblad torque. \citet{b20} showed that the corotation
torque becomes positive in a disc with the high opacity, reducing the
migration velocity of a planet. This process was recognized and
intensively studied by many
researchers \citep{b2, b21, b22, b17, b24, b34, b33, b45, b38}. The
positive corotation torque cancels
the negative Lindblad torque when the entropy distribution has steep
negative slope. It was also shown
that a planet in an adiabatic disc with the steeper negative slope
starts to migrate even outward.

The corotation torque significantly decreases in an adiabatic disc
after a few libration time of a planet because the entropy
distribution becomes uniform within the horseshoe region \citep{b48, b45}.
Without the sustained corotation torque, only the Lindblad
torque is eventually exerted on a planet in an adiabatic disc, leading to the
inward migration of a planet with Earth mass in $10^{5}$~yrs
\citep{b30, b25}. This is a serious problem in the core accretion
model of planet formation. The lifetime of a disc, $10^{6-7}$~yrs, is
much longer than the timescale of the Type~I migration, making the
survival of planets in a disc difficult. However, recent radial
velocity surveys of extrasolar planets show that a significant
fraction of solar-type stars may harbor close-in super-Earths
\citep{b16, b10, b11}. Population synthesis models have great
difficulties to reproduce the observed semimajor axis distribution of
extrasolar planets once the Type~I migration
is included. The reduction of the migration velocity of a planet is
required \citep{b1, b28}.

The viscosity of
a disc prevents the saturation of the corotation torque \citep{b48}.
\citet{b14} investigated the planet-disc
interactions in an optically thick accretion disc, taking into account
the viscous heating and the radiative cooling. They found the disc gas
reaches a steady state, of which the temperature distribution has the
very steep slope in a region where ice exists in a condensed form. The
corotation torque becomes
always positive in this region and the direction of the planet migration becomes
outward. \citet{b46}
comprehensively examined the various mechanisms to halt planet
migration in a disc around a young
star. They showed that the planet migration is halted at the ice line,
inside of which all the ice is
evaporated and solid particles are composed of rocks and metals. The
slope of the temperature
distribution becomes steep in the inner region. The positive
corotation torque becomes large enough to
make the total torque positive. On the other hand, the temperature
distribution beyond the ice line is
too shallow to cancel the negative Lindblad torque by the corotation
torque. A planet inside and outside
of the ice line is expected to move outward and inward, respectively.
They showed that the ice line is
the location to stop the planet migration in an optically thick accretion disc.

\citet{b19} showed that the torque is dependent on the magnitude of
the viscosity. The viscosity modifies the density of gas near a planet and decreases the torque density. \citet{b37} also showed that the corotation torque becomes very small when they consider a disc with high viscosity. In this study, we systematically examine the total torque
acting on a planet by an optically
thick accretion disc. It is not clear if the positive corotation
torque can always exceed the negative
Lindblad torque when we include viscosity in a disc. We make global
two-dimensional hydrodynamic
simulations to study the effect of the dissipation processes such as the viscosity and the radiation on the total torque.

This paper is organized as follows. In the section~2, we describe the
basic equations and a disc model we presume in this study. The dissipative
terms due to the viscosity and the radiation are included in the basic
equations. The temperature of the disc is determined by the energy
balance between the viscous heating and the radiative cooling. The
rate of the radiative cooling is dependent on the opacity of the disc,
which changes around the ice line. In the section~3, we show the
results of the two-dimensional hydrodynamic simulations. We show the
corotation torque decreases with an increase in the viscosity. The
total torque acting on the planet depends on the radiative cooling
rate and the viscous heating rate. We further derive the analytical
formula to relate the viscosity and the opacity when the total torque
exerted on the planet becomes zero. We summarize the results in the
section~4.

\section{Basic Equations and Disc Model}

\subsection{Basic Equations}

A planet excites density waves in a disc and changes
the density distribution of the disc. We examine the torque exerted on a
planet by an optically thick accretion disc. A planet with 5 Earth
masses rotates around a solar mass star in a fixed circular orbit. The
position vector of the planet from the star is denoted by $\textit
{\textbf {r}}_{\rm{p}}$. The problem is limited to a two-dimensional
flow, where all physical quantities (e.g., the surface density) depend
on $r$ and $\theta$, where $r$ is the distance from the star and
$\theta$ is the angle between the $x$-axis and the position vector.
Governing equations for the gas are the mass conservation, the
Navier-Stokes equations, and the energy equation with dissipative
terms due to the viscosity and the radiation.

We use a cylindrical coordinate where the star is located at the
center of the coordinate. The mass of the planet is much smaller than
that of the star and we neglect the indirect term. The mass
conservation equation and the Navier-Stokes equations read
\begin{equation}
\frac{\partial \Sigma}{\partial t}+\frac{1}{r}
\frac{\partial }{\partial r} \left(r \Sigma v_{r}\right)+\frac{1}{r}
\frac{\partial }{\partial \theta}\left(\Sigma v_{\theta}\right)=0,
\end{equation}
\begin{eqnarray}
\frac{\partial }{\partial t}\left(\Sigma v_{r}\right)+\frac{1}{r}
\frac{\partial }{\partial r} \left\{ r\left( \Sigma v_{r}^2 + p\right) \right\}
+\frac{1}{r} \frac{\partial }{\partial \theta} \left(\Sigma v_{r} v_{\theta}
\right) \nonumber \\
=\frac{\Sigma v_{\theta}^2}{r} + \frac{p}{r}
- \Sigma \frac{\partial \Phi}{\partial r} + f_{r},
\end{eqnarray}
\begin{eqnarray}
\frac{\partial }{\partial t}\left(\Sigma v_{\theta}\right)+\frac{1}{r}
\frac{\partial }{\partial r} \left( r \Sigma v_{r} v_{\theta}\right)
+\frac{1}{r} \frac{\partial }{\partial \theta} \left(\Sigma v_{\theta}^2 + p
\right) \nonumber \\
= -\frac{\Sigma v_{r} v_{\theta}}{r}
- \frac{\Sigma}{r}\frac{\partial \Phi}{\partial \theta} + f_{\theta},
\end{eqnarray}
where $\Sigma$ is the gas surface density, $p$ is the vertically
integrated pressure; $v_{r}$ and $v_{\theta}$ are the radial and
tangential velocities of the gas; $\Phi$ is the gravitational
potential. We consider a less massive disc and neglect the
self-gravity of the gas. The gravitational potential of the star and
the planet is given by
\begin{equation}
\Phi = -\frac{GM_{\odot}}{r} -
\frac{GM_{\rm{p}}}{\sqrt{r^2+r_{\rm{p}}^2- 2 r r_{\rm{p}} {\rm{cos}}
\psi + \epsilon^2 H_{\rm{p}}^2}},
\label{phi_eq}
\end{equation}
where $M_{\odot}$ is the mass of the sun, $M_{\rm{p}}$ is the mass of
the planet, $\psi$ is the angle between $\textit {\textbf {r}}$ and
$\textit {\textbf {r}}_{\rm{p}}$, and $H_{\rm{p}}$ is the scale height
of the disc at the location of the planet. The scale height is given
by
\begin{equation}
H_{\rm{p}} =
\frac{\sqrt{2} c_{\rm{p}} }{ \Omega_{\rm{p}} },
\label{scaleheight_eq}
\end{equation}
where $c_{\rm{p}}$ and $\Omega_{\rm{p}}$ are, respectively, the
isothermal sound velocity and the Keplerian angular velocity at
$r_{\rm{p}}$. The smoothing length parameter, $\epsilon$, is
introduced to include the effect of the scale height of a disc. It is noted that the small softening parameter leads to the
strong corotation torque \citep{b2, b24}. \citet{b38} examined the torque acting on a planet with several Earth
masses in an optically thin disc and found the torques calculated by
the simulations with $\epsilon = 0.3$ agree with that obtained by a
linear analysis. We adopt $\epsilon = 0.3$ in this study. The last
terms in the Navier-Stokes equations describe the radial and azimuthal
components of the viscous forces:
\begin{eqnarray}
f_{r} = \frac{1}{r}\frac{\partial}{\partial r} \left( r \sigma_{rr} \right) +
\frac{1}{r} \frac{\partial \sigma_{r \theta}}{\partial \theta} -
\frac{ \sigma_{\theta \theta}}{r}
\label{visco-fr}
\end{eqnarray}
and
\begin{eqnarray}
f_{\theta} = \frac{1}{r^2} \frac{\partial }{\partial r} \left(r^2
\sigma_{r \theta} \right) +\frac{1}{r} \frac{\partial \sigma_{\theta
\theta}}{\partial \theta},
\label{visco-ft}
\end{eqnarray}
where $\sigma_{rr}$, $\sigma_{r \theta}$, and $\sigma_{\theta \theta}$
are the components of the stress tensor and are, respectively, given
by
\begin{equation}
\sigma_{rr} = 2 \Sigma \nu \frac{\partial v_r }{\partial r},
\label{sigma-rr}
\end{equation}
\begin{equation}
 \sigma_{r \theta} = \Sigma \nu \left\{ r \frac{\partial }{\partial r}
\left( \frac{v_{\theta}}{r}\right) +\frac{1}{r} \frac{\partial
v_r}{\partial \theta} \right\},
\label{sigma-rt}
\end{equation}
and
\begin{equation}
\sigma_{\theta \theta} = \frac{2 \Sigma \nu}{r} \left( \frac{\partial
v_{\theta}}{\partial \theta} + v_r \right).
\label{sigma-tt}
\end{equation}
The gas is assumed to be ideal and the equation of state is given by
\begin{equation}
p=\frac{\Sigma k_{{\rm{B}}} T}{ \mu m_{\rm{H}}},
\label{EOS}
\end{equation}
where $k_{{\rm{B}}}$ is the Boltzmann constant, $T$ is the mid-plane
temperature, $m_{\rm{H}}$ is the mass of a hydrogen atom, and $\mu$ is
the mean molecular weight of the gas. We set up $\mu =$2.34.

The energy equation of the gas disc reads
\begin{eqnarray}
\frac{\partial (\Sigma e)}{\partial t}+\frac{1}{r}
\frac{\partial }{\partial r} \left\{ r v_{r} \left( \Sigma e + p
\right) \right\}
+\frac{1}{r} \frac{\partial }{\partial \theta} \left\{ v_{\theta} \left(
\Sigma e + p \right) \right\} \nonumber \\
= v_{r} \frac{\partial p}{\partial r}
+ \frac{v_{\theta}}{r}\frac{\partial p}{\partial \theta} + W - Q,
\label{BES}
\end{eqnarray}
where $e$ is the specific energy of the gas given with the pressure by
\begin{equation}
e = \frac{p}{(\gamma - 1) \Sigma} .
\label{E_eq}
\end{equation}
In Eq.($\ref{E_eq}$), $\gamma$ is the ratio of the specific heats at
constant pressure and volume. We use $\gamma= 4/3$ for a
two-dimensional disc as given by \citet{b15}. The energy of the gas
is increased by the viscous heating term. The viscous heating term,
$W$, is given by
\begin{eqnarray}
W = \sigma_{rr} \frac{\partial v_r }{\partial r} + \sigma_{r \theta}
\left\{ r \frac{\partial}{\partial r} \left( \frac{v_{\theta}}{r}
\right) + \frac{1}{r} \frac{\partial v_r}{\partial \theta} \right\} \nonumber \\
+ \sigma_{\theta \theta} \left( \frac{1}{r} \frac{\partial
v_{\theta}}{\partial \theta} + \frac{v_r}{r} \right).
\label{visco-heat}
\end{eqnarray}
 On the other hand, the energy of the gas is decreased by the
radiative cooling term of the disc, $Q$. The radiative cooling term,
$Q$, is given by
\begin{equation}
Q=2 \sigma_{\rm{SB}} T^4_{\rm{eff}},
\label{Qequation}
\end{equation}
where $\sigma_{\rm{SB}}$ is the Stefan-Boltzmann constant and
$T_{\rm{eff}}$ is an effective temperature of a disc. The radiative
flux in a radial direction is neglected because it is much smaller
than $Q$ by a factor of $H_{\rm{p}} / r_{\rm{p}}$. The effective
temperature is related to the mid-plane temperature of the disc as
\citep{b39}
\begin{equation}
T=\tau^{1/4}_{\rm{eff}} T_{\rm{eff}},
\label{TandTeff}
\end{equation}
with
\begin{equation}
\tau_{\rm{eff}} = \frac{3\tau}{8} + \frac{\sqrt{3}}{4} + \frac{1}{4\tau},
\label{taueffdef}
\end{equation}
where $\tau$ is the optical depth of a disc. The optical depth of a
disc is defined by
\begin{equation}
\tau = \frac{1}{2} \xi_{\rm{gr}} \kappa_{0} \Sigma,
\label{taudef}
\end{equation}
where $\kappa_{0}$ is the best fit function of the opacity \citep{b41}
and $\xi_{\rm{gr}}$ is the so-called grain content factor introduced
by \citet{b42}. Micron size dust particles are the main source of the
opacity. Collisions between particles might produce a large number of
small dust particles, increasing the opacity of a disc
\citep{b35}. The opacity might decrease due to capture of small dust
particles by large particles. It is valuable to study the
torque acting on a planet in a disc with various values of
$\xi_{\rm{gr}}$. We treat $\xi_{\rm{gr}}$ as a parameter of the
simulations and consider the discs with $0.1 \le \xi_{\rm{gr}} \le 100$.

The opacity strongly depends on the temperature of a disc \citep{b41}:
\begin{equation}
\kappa_{0} = \left \{
\begin{array}{lc}
5 \times 10^{-3} T & (210 {\rm{K}} \leq T < 2000{\rm{K}}), \\
2 \times 10^{16} T^{-7} & (170{\rm{K}} \leq T<210{\rm{K}}), \\
2 \times 10^{-4} T^2 & (T < 170{\rm{K}}).
\end{array}
\right.
\label{kappag}
\end{equation}
Equation~($\ref{kappag}$) was derived from the available opacity data,
which incorporates small particles found in interstellar medium. Dust
particles are composed of rocks and metals in a region of disc with $T
\geq 210~\mbox{K}$, while ice is added to form dust particles in a
region of disc with $T < 170\mbox{K}$. In the transition region
between the cold and the hot region, the ice is evaporating. We divide
a disc into three regions: the region~1 with $T \geq 210~\mbox{K}$,
the region~2 with $170~\mbox{K} \leq T < 210~\mbox{K}$, and the region~3
with $T < 170~\mbox{K}$.

For the later convenience, we write all the quantities in a
non-dimensional form using the unit length $r_0$= 1~AU, the unit mass
$M_{\odot}$, and the unit time $\Omega_{\rm{0}}^{-1}$ where
$\Omega_{\rm{0}}$ is the Kepler frequency at 1~AU. The normalized
quantities are denoted by a tilde (e.g., $\tilde{r}_{\rm{p}}$).

\subsection{Disc Model}

\citet{b13} obtained images of protoplanetary discs around T Tauri
stars in Taurus using the thermal emission of dust. They found that
the surface density distributions of protoplanetary discs are
described by the power-law distribution with the power-law index of
0-1. The surface densities of the discs at 100~AU are in a range
between 0.1 and 10~${\rm{g/cm^{2}}}$. In this study we adopt a disc
model, of which the initial condition is given by the power-law distribution:
\begin{equation}
\Sigma_{\rm{ini}}= \Sigma_0  \tilde{r}^{-\alpha},
\label{sigeq}
\end{equation}
where $\Sigma_0$ is the surface density at $r_0$ and is set to be
100~$\rm{g/cm^2}$.

\citet{b13} also found disc radii expand with time, suggesting that
most of the gas migrates toward inward a central star by transporting
the angular momentum outward. We adopt an accretion disc model
developed by \citet{b43}, in which the accretion rate of gas is
constant throughout a disc. The viscosity $\nu$ is expressed by a
power-law function of the distance from the central star to satisfy
the constant accretion rate of a disc as
\begin{equation}
\nu = \xi_{\rm{v}} \nu_0 \tilde{r}^{\alpha},
\label{viscocoeff}
\end{equation}
where $\nu_0$ is the kinematic viscosity at $r_0$ and we set $\nu_0 =
4.2 \times 10^{15} {\rm{cm^2/s}}$ and $\xi_{\rm{v}}$ is a viscous
strength factor. The value of $\nu_0$ is approximately
equivalent to 0.05 in terms of the alpha coefficient of \citet{b49}.
The accretion rate of a disc with the
kinematic viscosity, $\nu_0$, and the surface density given by
Eq.~($\ref{sigeq}$) becomes $6.3 \times
10^{-8} M_{\odot} /{\rm{yr}}$. Observations of discs suggest that the
accretion rates of discs are $\sim
1 \times 10^{-8} M_{\odot} /{\rm{yr}}$ \citep{b7, b13}. We treat
$\xi_{\rm{v}}$ as another parameter in
addition to $\xi_{\rm{gr}}$ to consider discs with various accretion
rates of $0.1 \le \xi_{\rm{v}} \le 1$.
Two parameters, $\xi_{\rm{gr}}$ and $\xi_{\rm{v}}$, determine the
structure of a disc.

We derive the temperature distribution of an optically thick accretion
disc following \citet{b14}. The temperature distribution of the disc
is determined by the balance between the viscous heating term, $W$,
and the radiative cooling term, $Q$. Assuming that the viscous heating
is due to the Keplerian motion, we have $W$ given by
\begin{equation}
W = \frac{9}{4} \Sigma \nu \Omega^2.
\label{Wapp}
\end{equation}
Applying Eqs.~($\ref{TandTeff}$)-($\ref{kappag}$) to $Q$ with the
assumption of $\tau \gg 1$, we obtain the temperature distribution of
the disc
from $W=Q$ as
\begin{equation}
T = \left \{
\begin{array}{lc}
210 \left( \frac{r}{r_{12}} \right)^{-(\alpha+3)/3} & (210{\rm{K}}
\leq T < 2000{\rm{K}}), \\
210 \left( \frac{r}{r_{12}} \right)^{-(\alpha+3)/11} &
(170{\rm{K}} \leq T<210{\rm{K}}), \\
170 \left( \frac{r}{r_{23}} \right)^{-(\alpha+3)/2}  & (T < 170{\rm{K}}),
\end{array}
\right.
\label{teq}
\end{equation}
where $r_{12}$ and $r_{23}$ are the distances from the star to the
boundary of the regions~1 and 2 and that of the regions~2 and 3,
respectively. For later convenience, the power-law indexes of the
temperature distribution in the $i$-region are denoted by $-\beta_i$
(e.g., $\beta_1 = (\alpha + 3)/3$).

Figure~$\ref{Initial-state-distribution}$ shows the surface density
distribution and the temperature distribution of the disc with $\alpha
= 1.0$. It is seen from this figure that the temperature distribution
has three different slopes. The boundary between the regions~1 and 2
and that between the regions~2 and 3 are located at 1.4~AU and 2.5~AU,
respectively. The temperature distribution is nearly flat in the
region~2 ($\beta_2 =4/11$). The power-law indexes of the temperature
distributions are given by $\beta_1 = 4/3$ in the region~1 and
$\beta_3 =2.0$ in the region~3.

The larger viscosity and/or opacity increase the temperature of the
disc. The boundary positions move outward with increases in the
viscosity and/or opacity. The location of the boundary is expressed by
the two parameters, $\xi_{\rm{gr}}$ and $\xi_{\rm{v}}$,  and the
boundary position between the regions~1 and 2 is given by
\begin{equation}
\tilde{r}_{12} = 1.4 \left( \frac{\xi_{\rm{v}} \xi_{\rm{gr}}}{1.0}
\right)^{1/4} .
\label{r12def}
\end{equation}
The boundary position between the regions~2 and 3 is given by
$\tilde{r}_{23} = 1.8 \tilde{r}_{12}$ in the case of $\alpha = 1.0$.
This boundary positions were derived by \citet{b46} as well. It is noted that 
the boundary position also depends on the surface density. We plot the boundary positions of the
regions~1 and 2, $\tilde{r}_{12}$, as a function of $\xi_{\rm{gr}}$
and $\xi_{\rm{v}}$ in Fig.$\ref{r12contour}$. The solid, dashed,
dot-dashed, and dotted curves correspond to $\tilde{r}_{12} = 2.4$,
1.4, 1.2, and 0.8, respectively. It is found that the boundary
position between the regions~1 and 2 ranges from 0.5~AU to 2.5~AU. The
filled circles represent the parameter sets $(\xi_{\rm{gr}},
\xi_{\rm{v}})$ used in our numerical simulations.

\section{Numerical Method and Simulation Results}

A planet generates density waves in a disc inside and outside of the
orbit of the planet. The inner and outer density waves exert the
positive and negative torques on the planet, respectively. The gravity
of an outer density wave is a little stronger than that of an inner
density wave because an outer density wave is closer to the planet due
to the pressure gradient of a disc, leading to the negative Lindblad
torque \citep{b30}. Another torque acts on the planet by a gas element
in the horseshoe orbit of the planet \citep{b2,b22}. When a gas
element in the horseshoe orbit approaches a planet, the angular
momentum is exchanged between the gas element and the planet. This is
called the corotation torque.

The entropy of the gas is expressed as $S = p/\Sigma^{\gamma} \propto
r^{\lambda}$, where $\lambda
= (\gamma-1)\alpha - \beta_i$. The corotation torque
is dependent on the entropy
distribution of a disc. It was shown that the corotation torque
consists of linear and non-linear corotation
torque. The non-linear corotation torque comes from the outgoing
boundaries of the horseshoe region, that is the separatrix
\citep{b48, b24}. In a non-barotropic disc, the vortensity is changed
after the horseshoe
U-turn. The enhanced corotation torque is not due to the adiabatic
compression, but comes from a
singular streamline at the separatrix. Since the change of the
vortensity is proportional to the radial
entropy gradient of a disc, the large corotation torque is induced in
a disc with the large magnitude of
$\lambda$. \citet{b38} showed that the positive corotation torque is
comparable with the negative
Lindblad torque when $\lambda = -0.4$ in an adiabatic disc. The total
torque acting on a planet is
determined by the sum of the negative Lindblad torque and the corotation torque.

The disc we consider in the section~2.2 has the surface density
distribution with $\alpha = 1.0$ and the temperature distribution with
$\beta_1 = 4/3$ in the region~1, $\beta_2 = 4/11$ in the region~2, and
$\beta_3 = 2.0$ in the region~3. The power-law indexes of the entropy
distribution are calculated as $\lambda = -1.0$ in the region~1,
$\lambda = -3.0 \times 10^{-2}$ in the region~2, and $\lambda = -1.7$ in the
region~3. \citet{b38} found that the total torque acting on a planet
in an adiabatic disc becomes positive when $\lambda < -0.4$. We apply
this to the optically thick disc and predict that a planet migrates
outward in the regions~1 and 3 while it moves inward toward the star in the
region~2. Planets in the regions~1 and 2 are expected to move toward the
boundary of the regions~1 and 2. This might help forming larger
planets. However, it is not clear whether a planet still moves outward
in the region~1 even if dissipative processes are included in a disc.
We make a number of numerical simulations to examine the torque acting
on a planet in a disc with dissipative processes due to the viscosity
and the radiation.

\subsection{Numerical Method}

We use two-dimensional equidistant grids in $r$ and $\theta$ with a
resolution of ($N_r$, $N_{\rm{\theta}}$) = (640, 3072). The inner and
outer radii of the disc are given by $r_{\rm{min}}/r_{\rm{p}}=$0.4 and
$r_{\rm{max}}/r_{\rm{p}}= 2.0$, respectively. The damping boundary conditions
\citep{b4}, where all components are relaxed towards their initial
state, are used in order to reduce wave reflection from these
boundaries. All the quantities in the inner and outer boundaries are
always fixed to be the initial values.

We develop a two-dimensional global hydrodynamic computer program with the
gravitational forces of a star and a planet and a dissipative term
\citep{b38}. The basic equations are solved simultaneously using the
finite volume method with an operator splitting procedure. The source
terms, which include the gravity, the viscous
force, and the radiation, are computed with a second order Runge-Kutta
scheme. The advection terms are calculated with a second order
MUSCL-Hancock scheme and an exact Riemann solver \citep{b26, b12}.

The angular momentum of the disc gas is transferred to the planet. The
transfer rate of the angular momentum from the disc gas at $r$ to the
planet is given by
\begin{equation}
\Gamma_{r} = \int_0^{2 \pi} \Sigma
\frac{\partial  \Phi}{\partial \theta} r{\rm{d}}\theta .
\label{Gamma_den_def}
\end{equation}
By integrating the torque density over the radial distance, we obtain
the total torque acting on the planet from the disc:
\begin{equation}
\Gamma = \int_{r_{\rm{min}}}^{r_{\rm{max}}} \Gamma_{r} {\rm{d}}r.
\label{Gamma_def}
\end{equation}
The torque and the torque density are normalized by $\Gamma_0$ and
$\Gamma_0/r_{\rm{p}}$, where $\Gamma_0$ is $(M_{\rm{p}}/M_{\odot})^2
(r_{\rm{p}}/H_{\rm{p}})^2 \Sigma_{\rm{p}} r_{\rm{p}}^4
\Omega_{\rm{p}}^2$.

\subsection{Simulation Results}

\subsubsection{The torque density acting on a planet in an optically
thick accretion disc}

We consider two adiabatic discs that have the temperature
distributions with the power-law indexes $4/3$ and $4/11$, which
correspond to that of the region~1 and the region~2 of the optically thick disc,
respectively. The power-law index of the temperature distribution is
expressed as $-\beta$. Both discs have the initial surface density
with $\alpha = 1.0$. Hereafter we call the adiabatic disc with
$(\alpha, \beta)=(1.0, 4/3)$ and that with $(\alpha, \beta)=(1.0, 4/11)$ the
disc~A and the disc~B, respectively. The power-law index of the entropy
distribution of the disc~A is $\lambda=-1.0$ and that of the disc~B is
$\lambda=-3.0 \times 10^{-2}$. \citet{b38} found that the corotation torque
increases with a decrease in the power index of the entropy
distribution in an adiabatic disc. Figure~$\ref{rp_adiabatic-lt25}$
shows the radial distribution of the torque density exerted on the
planet in the adiabatic discs at $t = 25 t_{\rm{p}}$, where
$t_{\rm{p}}$ is the rotational period of the planet. The planet is
located at 1~AU in the disc~A and 2~AU in the disc~B. The large
corotation torque is exerted on the planet in the disc~A due to the large entropy gradient, making the total torque positive. On the other hand,
the corotation torque in the disc~B is too weak to cancel the negative
Lindblad torque. The normalized total torques become 3.0 and $-2.4$ in
the disc~A and the disc~B, respectively.

The half width of the horseshoe region, $\delta x_{\rm{s}}$, is
estimated as \citep{b36, b23, b24}

\begin{equation}
\frac{\delta x_{\rm{s}}}{r_{\rm{p}}} = 2.1 \times 10^{-3} \sqrt{\left(
\frac{M_{\rm{p}}}{M_{\rm{E}}} \right) \left(
\frac{r_{\rm{p}}}{H_{\rm{p}}} \right) },
\label{xs}
\end{equation}
where $M_{\rm{E}}$ is the Earth mass and $H_{\rm{p}}$ is the scale
height of the disc at $\tilde{r} = \tilde{r}_{\rm{p}}$:
\begin{equation}
\frac{H_{\rm{p}}}{r_{\rm{p}}}= 4.8 \times 10^{-2} \left(
\frac{\tilde{r}_{\rm{p}}}{1.4} \right)^{(1 - \beta)/2}.
\label{scaleHp_rp}
\end{equation}
\citet{b36} showed that the half width of the
horseshoe region of a planet scales with $M_{\rm{p}}^{1/2}$ as long as
the flow around a planet remains linear. Figure $\ref{horseshoewidth}$
shows the half width of the horseshoe region given by the numerical
simulations (the filled circles) together with the half width of the
horseshoe region given by Eq.($\ref{xs}$) (the curve). Both half
widths agree well except the large planet mass. \citet{b36} found the
planet mass when the flow linearity breaks in a 2D simulation. The
non-linearity of the flow around a planet becomes clear when the mass
of the planet becomes larger than $12M_{\rm{E}}
(H_{\rm{p}}/0.05r_{\rm{p}})^3$. The numerical results also start to diverge
from the half width given by Eq.($\ref{xs}$) when the mass of the
planet becomes larger than 10~Earth masses.

The ratios of the magnitude of the corotation torque to that of the
Lindblad torque are 1.6 and 0.6 in the disc~A and the disc~B,
respectively. Nearly the same Lindblad torques act on the planet in
both discs. The corotation torque determines the magnitude and sign of
the total torque exerted on the planet.

\begin{figure*}
\centerline{\includegraphics[scale=0.6,clip]{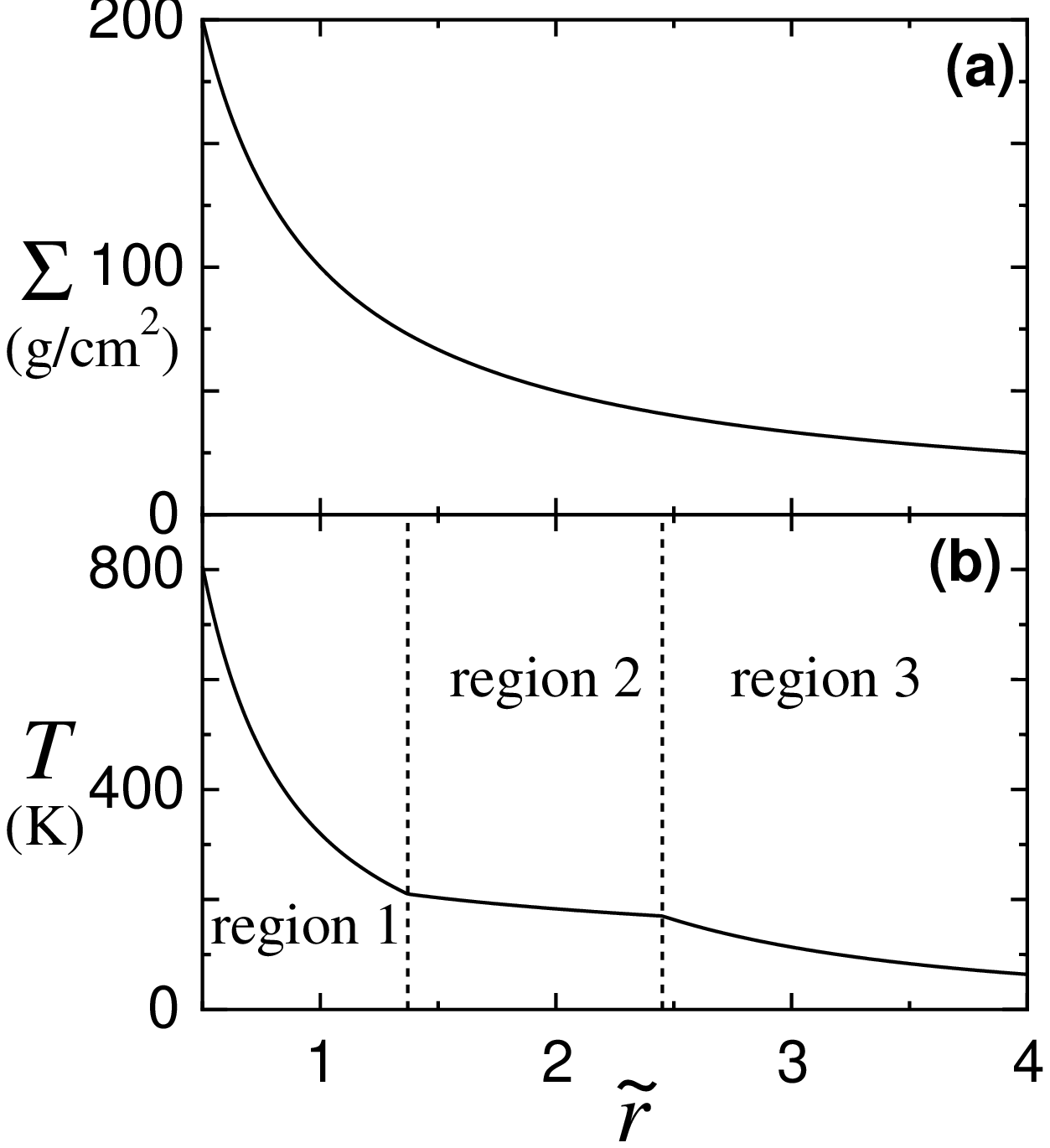}}
\caption{
(a) the surface density distribution and (b) the temperature
distribution of the disc with $\alpha = 1.0$. The
temperature distribution of the disc is determined by the balance
between the viscous heating and the radiative cooling. The temperature
distribution is described by the
power-law distributions with three different gradients. The power-law
index of the temperature distribution strongly depends on the opacity
of the disc. The dependence of the opacity on the temperature changes
due to the evaporation of ice. The power-law indexes of
the temperature distribution change at 1.4~AU and 2.5~AU.
}
\label{Initial-state-distribution}
\end{figure*}

\begin{figure*}
\centerline{\includegraphics[scale=0.6,clip]{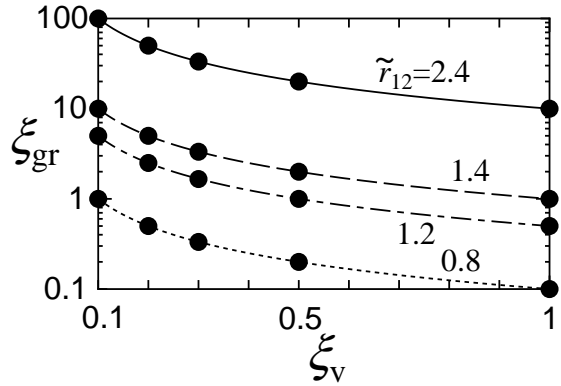}}
\caption{
The boundary position between the regions~1 and 2, $\tilde{r}_{12}$,
as a function of $\xi_{\rm{gr}}$ and $\xi_{\rm{v}}$. The solid,
dashed, dot-dashed, and dotted lines correspond to $\tilde{r}_{12} =
2.4$, 1.4, 1.2, and 0.8, respectively. We use disc models with 20
parameter sets $(\xi_{\rm{gr}}, \xi_{\rm{v}})$ marked by the filled
circles.
}
\label{r12contour}
\end{figure*}

\begin{figure*}
\centerline{\includegraphics[scale=0.6,clip]{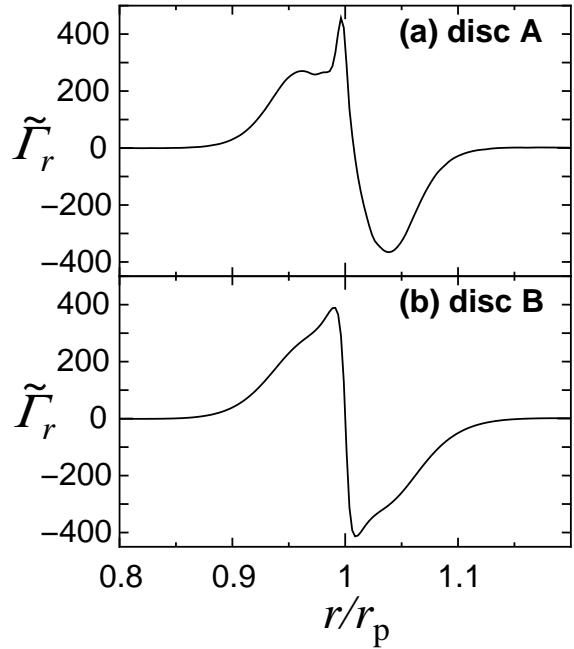}}
\caption{
The radial distributions of the torque density acting on the planet
with 5~Earth masses by the adiabatic disc at $t = 25 t_{\rm{p}}$. The
power-law indexes of the temperature distributions of the disc are
$\beta = 4/3$ and $\beta = 4/11$ in the panels (a) and (b),
respectively. The planets are located at 1~AU and 2~AU in the panels (a)
and (b), respectively. The horizontal axis is normalized by $r_{\rm{p}}$. 
The strong corotation torque is found in the panel~(a), leading to the outward
migration of the planet.
}
\label{rp_adiabatic-lt25}
\end{figure*}

\begin{figure*}
\centerline{\includegraphics[scale=0.6,clip]{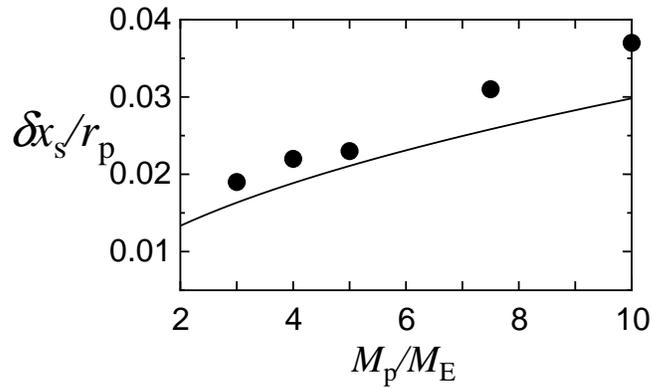}}
\caption{The half width of the horseshoe region as a function of the
planet mass. The horizontal axis is normalized by the Earth mass. The
optically thick disc with the accretion rate of $\xi_{\rm{v}} = 0.1$
is used. The numerical half width of the horseshoe region (filled circles) is 
evaluated at $t = 25 t_{\rm{p}}$. The planet is located at
$\tilde{r}_{\rm{p}} = 1.0$. The curve is drawn using the analytical
expression for the half width of the horseshoe region given by
\citet{b24}.
}
\label{horseshoewidth}
\end{figure*}

\begin{figure*}
\centerline{\includegraphics[scale=0.6,clip]{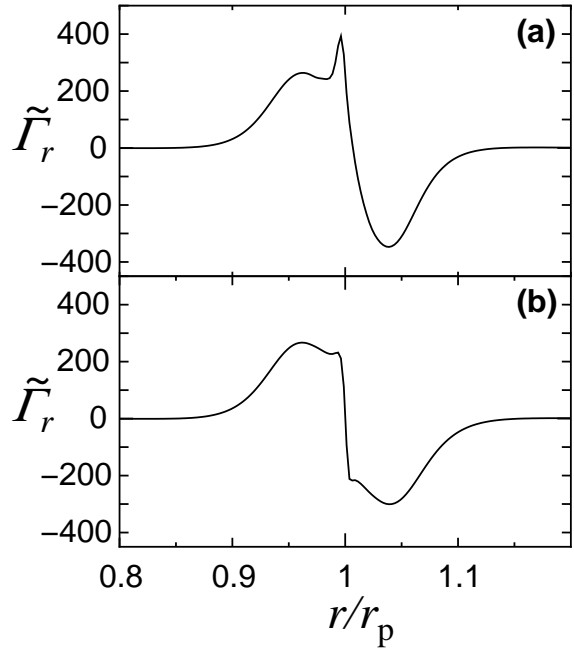}}
\caption{
Same as Fig.~3, but the optically thick accretion disc with
$(\xi_{\rm{gr}}, \xi_{\rm{v}}) =$ (10, 0.1). The normalized total
torque acting on the planet is 2.1 in the panel~(a), while it becomes
$-1.5$ in the panel~(b). The planet migrates outward in the region~1 and
inward in the region~2. Planets might accumulate at the boundary of
the regions.
}
\label{dmdt1e-9kappa10_lt25}
\end{figure*}

We perform the simulation to find the torque density exerted on the
planet in the optically thick accretion disc with $(\xi_{\rm{gr}},
\xi_{\rm{v}}) = (10, 0.1)$. The temperature distribution is the same
as that of Fig.~$\ref{Initial-state-distribution}$. The boundary
position between the regions~1 and 2 is 1.4~AU and that between the
regions~2 and 3 is 2.5~AU. Figure~$\ref{dmdt1e-9kappa10_lt25}$ shows
the radial distributions of the torque density acting on the planets
in the region~1 (panel (a)) and in the region~2 (panel (b)) at $t = 25
t_{\rm{p}}$. The planets are located at 1~AU in the region~1 and 2~AU
in the region~ 2. The radial distribution of the torque density in the
region~1 is similar to that in the disc~A. The torque density has a
local maximum and a local minimum at $|1- r / r_{\rm{p}}|= 3.3 \times
10^{-2}$. The distances from the planet to the local maximum and
minimum can be approximated by the scale height of the disc in the
region~1, indicating that this torque corresponds to the
Lindblad torque.

\begin{figure*}
\centerline{\includegraphics[scale=0.6,clip]{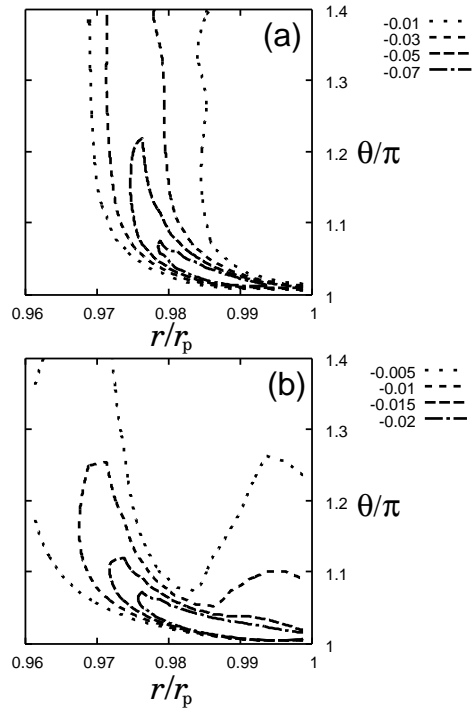}}

\caption{ Contour of the vortensity in (a) the inviscid adiabatic disc (disc~A) and (b) the accretion disc with the non-vanishing viscosity (($\xi_{\rm{v}}, \xi_{\rm{gr}}$)=(0.1,10)) at $t
= 10 t_{\rm{p}}$. The initial distribution of vortensity is subtracted
from the vortensity for better contrast. The horizontal and
vertical axes correspond to $r/r_{\rm{p}}$ and $\theta / \pi$,
respectively. The planet with 5~Earth masses is located at 1~AU.}
\label{vortensitycontour}
\end{figure*}

The flow pattern within the horseshoe region is modified by the viscosity and the vortensity is no longer conserved during the U-turn. The vortensity is radially transferred by the viscosity. We compare the vortensity distributions of the inviscid disc, that is the adiabatic disc, and the accretion disc with non-vanishing viscosity. Figure $\ref{vortensitycontour}$ shows the contour of the
vortensity within the horseshoe region of the planet. It is seen from
Fig.~$\ref{vortensitycontour}$ that the location of the minimum
vortensity moves inward toward the central star when the viscosity is
included. The viscosity plays an important role to diffuse the
vortensity. In a disc with the low viscosity, the corotation torque is dominated
by the non-linear corotation torque. The non-linear corotation torque
decreases with the increasing viscosity.

\begin{figure*}
\centerline{\includegraphics[scale=0.6,clip]{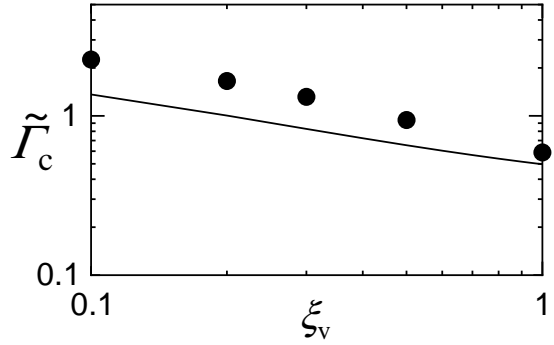}}
\caption{
The corotation torques acting on the planet with 5 Earth masses given by
the analytical expression of \citet{b48} (solid line) and the numerical
results (filled circles). The horizontal and vertical lines represent
the viscous strength factor, $\xi_{\rm{v}}$, and the normalized
corotation torque, $\tilde{\Gamma}_{\rm{c}}$, respectively. In
deriving the numerical corotation
torque, we used Eq.($\ref{xs}$) for the half width of the horseshoe region.
The planet is located at 1~AU.
}
\label{zerothermalcorotorq}
\end{figure*}

The radiation restores the entropy distribution of a
disc and is important to prevent the saturation of the corotation
torque. \citet{b48} derived the analytical expression of the
corotation torque acting on a planet by a disc with viscosity and
thermal diffusivity. In this study, the cooling due to the radiation
is compensated by the viscous heating. The radiation only works to
maintain the thermal structure of a disc. Our numerical results are
compared with the analytical expression of the corotation torque given by
\citet{b48} in Fig.~$\ref{zerothermalcorotorq}$. We remove the
contribution of thermal diffusivity in the analytical expression. Our
results agree well with the analytical result. In this study, the
radiation affects the corotation torque exerted on a planet through
the disc temperature profile. However, it is important to note that
radiation significantly influences the corotation torque as pointed by
\citet{b48}.

The timescale for the gas to spread across the horseshoe width by the
viscosity is defined by
\begin{equation}
\tau_{\rm{visc}} = \frac{\delta x_{\rm{s}}^2}{3 \nu}.
\label{visco_time}
\end{equation}
We substitute Eqs.~($\ref{viscocoeff}$) and ($\ref{xs}$) into
Eq.~($\ref{visco_time}$) to estimate the viscous timescale for the
disc with $\xi_{\rm{v}} = 0.1$ at $r_{\rm{p}} = $ 1~AU:
\begin{equation}
\tau_{\rm{visc}} = 16 \left( \frac{\xi_{\rm{v}}}{0.1} \right)^{-1}
\left(\frac{H_{\rm{p}}/r_{\rm{p}}}{0.05} \right)^{-1}
\left( \frac{M_{\rm{p}}}{5M_{\rm{E}}} \right )
\Omega_0^{-1}.
\label{visco_estimate}
\end{equation}
We consider the pressure effect of the gas disc and obtain the turnover time
along the horseshoe orbit in front of the planet at
$r_{\rm{p}} = $ 1~AU \citep{b2}:
\begin{equation}
\tau_{\rm{turn}} = 12 \left(\frac{H_{\rm{p}}/r_{\rm{p}}}{0.05} \right)^{3/2}
\left( \frac{M_{\rm{p}}}{5M_{\rm{E}}} \right )^{-1/2} \Omega_0^{-1}.
\label{turn_time}
\end{equation}

The viscous timescale is nearly equal to the turnover time in the
disc with $\xi_{\rm{v}} = 0.1$. The effect of the viscosity on the corotation torque then starts to be important. Note that the cooling timescale by the radiation is the same with the viscous timescale in the disc.

The torque density near the orbit of the planet in the region~2 is
smaller than that in the disc~B. \citet{b18} used a
local shearing box calculation and studied the gravitational
interactions between a planet and an optically thin disc, taking into
account of energy dissipation by radiation. They found that the oval
shape of the
density contour profile near a planet is tilted with respect to the
direction toward the central star. The asymmetry of the
density structure increases the one-side torque. They showed that the
density contour is less inclined with the decreasing opacity. The
adiabatic
disc can be considered to have much larger opacity than
the region~2, yielding the larger torque density near the
planet.

\citet{b45} have derived the torque formula for the Lindblad torque and the
corotation torque exerted on a planet by a disc with viscosity and
thermal diffusion. We find that the corotation and Lindblad torques
obtained in our simulation agree with those calculated from their
torque formula within 30~\%.
The total torque is given by the sum of the Lindblad torque and the
corotation torque. The total torque becomes $\tilde{\Gamma} = 2.1$ in
the region~1 and it is slightly smaller than that in the disc~A,
$\tilde{\Gamma} = 3.0$. The density enhancement is reduced as the
vortensity smears by the viscosity. However, the viscosity is too small to
remove the corotation torque. Planets accumulate at the ice line as suggested
by \citet{b46}.

\begin{figure*}
\centerline{\includegraphics[scale=0.6,clip]{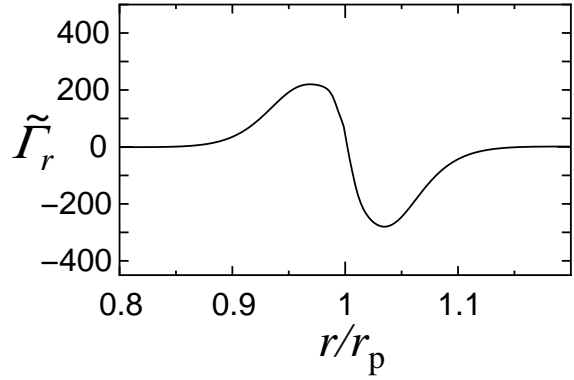}}
\caption{
The radial distribution of the torque density acting on the planet with 
5 Earth masses by the region 1 of the disc with $(\xi_{\rm{gr}}, \xi_{\rm{v}}) =$ (1.0, 1.0). The planet is located at 1~AU. The total torque acting on the planet becomes negative, $\tilde{\Gamma}=-2.4$. Planets move inward toward the central star even in the region~1.
}
\label{dmdt1e-8kappa1_lt25}
\end{figure*}

We further examine the effect of the viscosity of a disc on the torque
density exerted on the planet, using the large value of the viscosity,
$\xi_{\rm{v}} = 1.0$. We utilize the same surface density distribution and
the temperature distribution given by
Fig.~$\ref{Initial-state-distribution}$.
Figure~$\ref{dmdt1e-8kappa1_lt25}$ shows the radial distribution of
the torque density exerted on the planet in the region~1 at
$t = 25 t_{\rm{p}}$. The corotation torque is lost because the viscosity is
large enough to smear the density enhancement. In this case, the
viscous timescale becomes much shorter than the turnover time (see
Eqs.($\ref{visco_estimate}$) and ($\ref{turn_time}$)). The total
torque becomes negative ($\tilde{\Gamma} = -2.4$), leading to the
inward migration of the planet in the region~1.

\begin{figure*}
\centerline{\includegraphics[scale=0.6,clip]{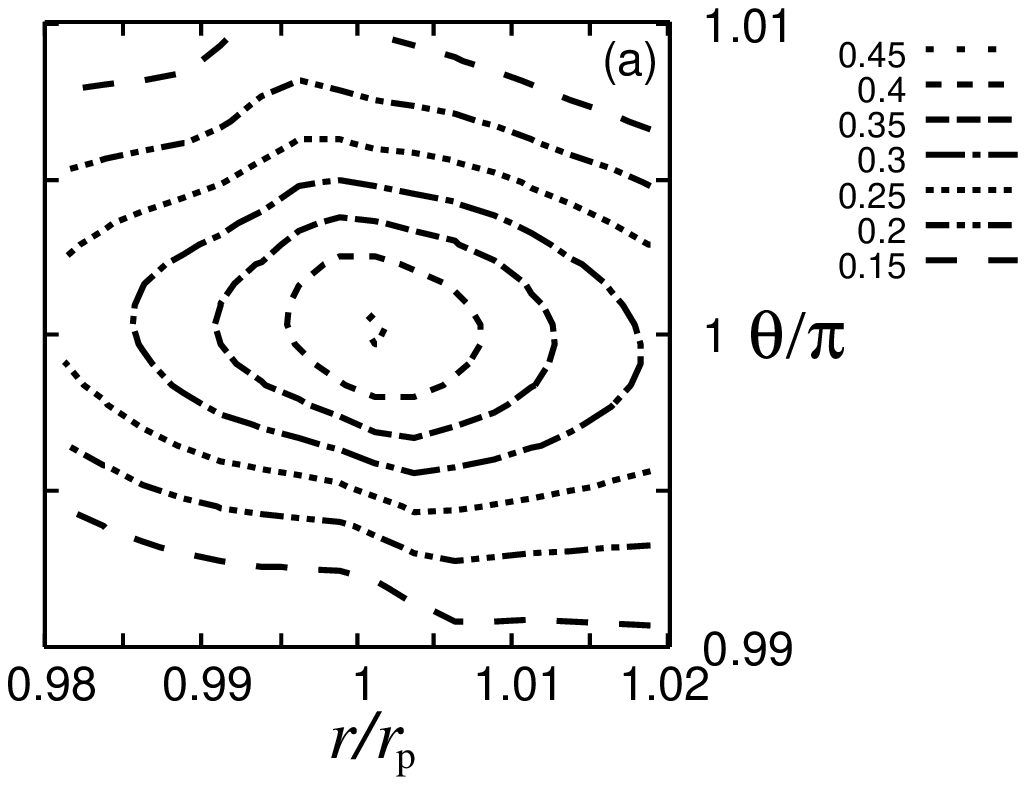}}
\centerline{\includegraphics[scale=0.6,clip]{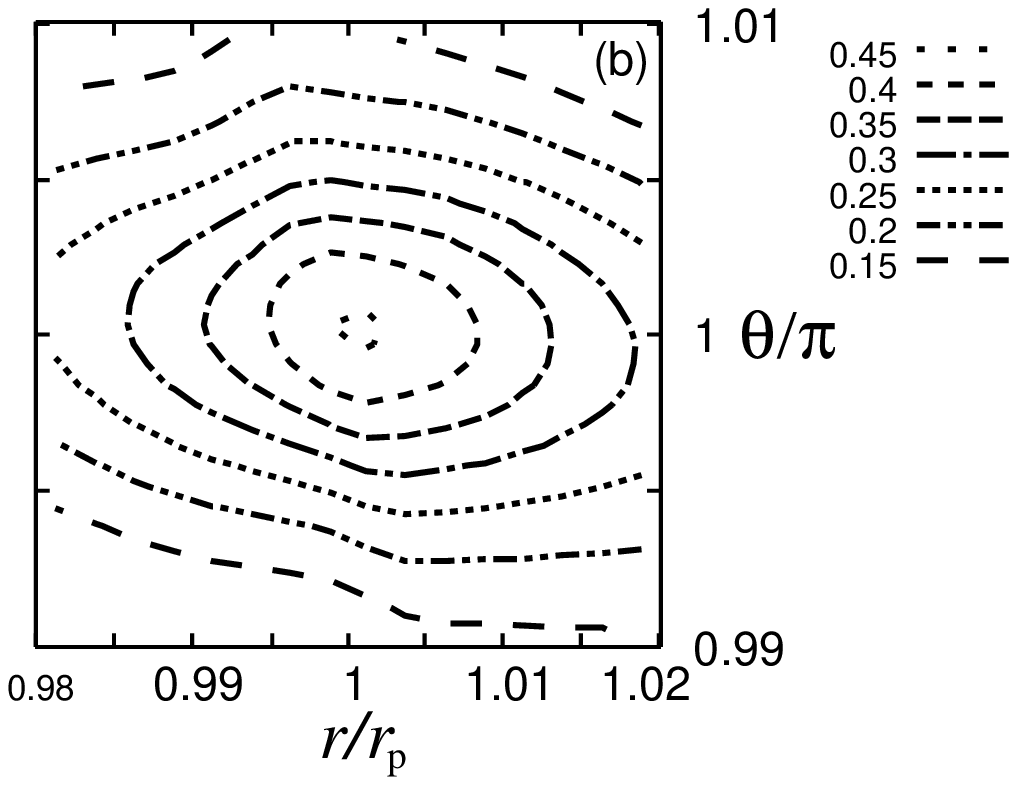}}
\centerline{\includegraphics[scale=0.6,clip]{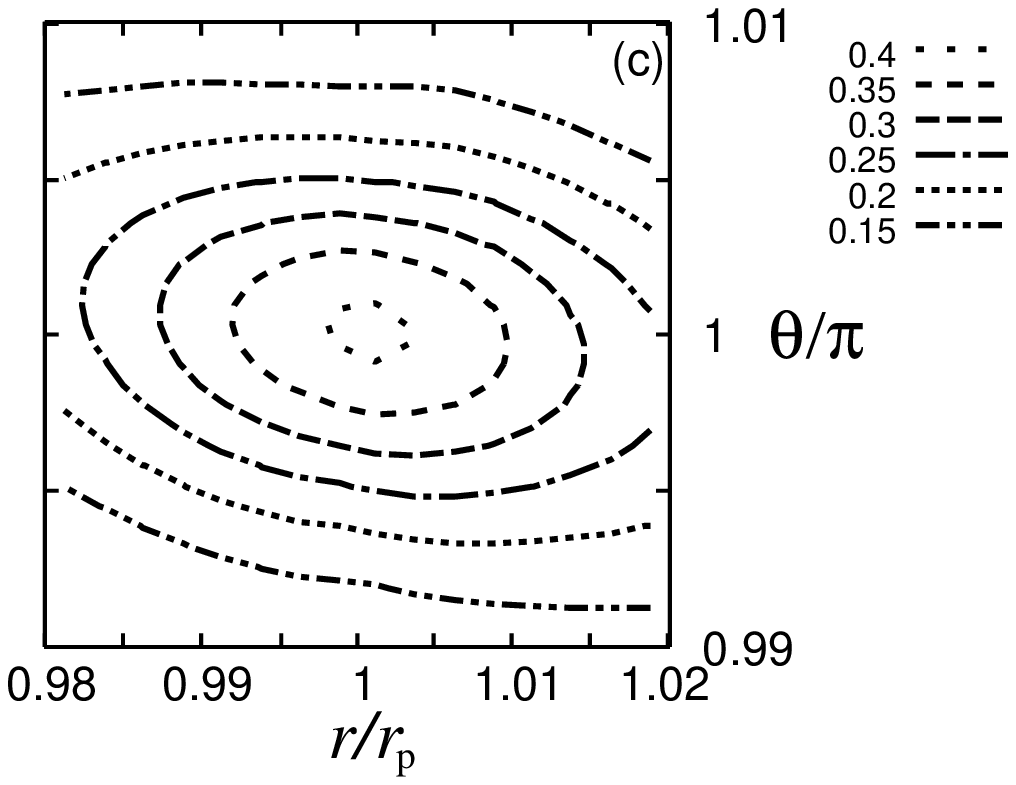}}
\caption{
The contour plots of the surface density enhancement
$\Sigma(t= 25 t_{\rm{p}})/ \Sigma_{\rm{ini}} - 1.0$ around the
planet with 5~Earth masses (a) in an adiabatic disc (disc~A), (b) in
the region~1 of the disc with $(\xi_{\rm{gr}}, \xi_{\rm{v}}) =$ (10,
0.1), and (c) in the region~1 of the disc $(\xi_{\rm{gr}},
\xi_{\rm{v}}) =$ (1.0, 1.0), respectively. The horizontal axis is
normalized by $r_{\rm{p}}$ and the vertical axis is divided by $\pi$.
The planet is located at $(r, \theta)=(r_{\rm{p}}, \pi)$.
The density contour in the panel~(b) is similar to that in the panel~(a)
because the viscosity is too small to
decrease the density enhancement. On the other hand, the density
enhancement is greatly reduced by the viscosity in the panel~(c).
}
\label{contourplot_density}
\end{figure*}

The density contours around the planet are shown to make clear the
effect of the viscosity on the torque density.
Figure~$\ref{contourplot_density}$ shows the contours of the density
fluctuations, $\Sigma(t= 25 t_{\rm{p}}) / \Sigma_{\rm{ini}} -1.0$, around the planet (a) in the disc~A, (b) in the region~1 of the
disc with $(\xi_{\rm{gr}}, \xi_{\rm{v}})=(10, 0.1)$, and (c) in the
region~1 of the disc with $(\xi_{\rm{gr}}, \xi_{\rm{v}})=(1.0, 1.0)$. The
enhancement of the density in the horseshoe orbit in the panel~(b) is
similar to that in the panel~(a). On the other hand, the density in
the horseshoe region is considerably reduced in the panel~(c) because
of the larger viscosity. The corotation torque decreases because the
viscous timescale is much shorter than the turnover time in the case of
$\xi_{\rm{v}} = 1.0$. Moreover, the inner and outer spiral density waves
are damped by the viscosity as well. This results in the small Lindblad torque.

\subsubsection{Dependence of the total torque on the opacity and the viscosity
of a disc}

\begin{figure*}
\centerline{\includegraphics[scale=0.6,clip]{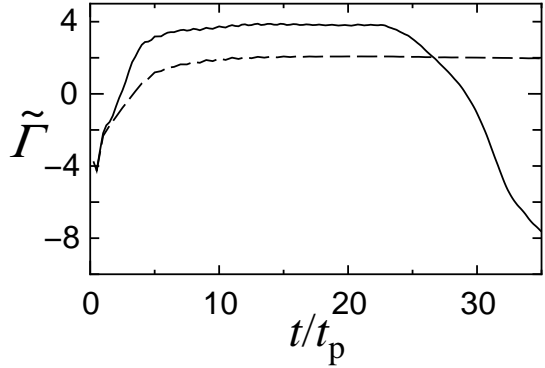}}
\caption{
The time evolution of the total torque acting on the planet with
5~Earth masses in the adiabatic disc (solid curve) and in the region~1
of the disc with $(\xi_{\rm{gr}}, \xi_{\rm{v}}) =$ (10, 0.1) (dashed
curve). In both cases, the planet is located at 1~AU. The total torque
decreases around $t = 25 t_{\rm{p}}$ in the adiabatic disc due to the
saturation of the corotation torque. On the other hand, the positive
total torque is maintained in the disc with $(\xi_{\rm{gr}},
\xi_{\rm{v}}) =$ (10, 0.1).
}
\label{totaltorqueevolution}
\end{figure*}

The corotation torque decreases in an adiabatic disc because the
entropy of the gas in the horseshoe orbit of the planet tends to
become uniform after the synodic period of the planet \citep{b2, b22,
b37}. This is called the saturation of the corotation torque and
happens in an adiabatic disc. The corotation torque does not saturate
in a viscous disc because the vortensity is transferred from the
horseshoe region to the outer region of the disc by viscosity
\citep{b22, b37}.
Figure~$\ref{totaltorqueevolution}$ shows the time evolutions of the
total torque acting on the planet in the adiabatic disc (the disc~A)
and in the optically thick accretion disc with $(\xi_{\rm{gr}},
\xi_{\rm{v}}) = (10, 0.1)$ (the region~1), which are represented by
the solid and dotted curves, respectively. The total torque increases
with time at the beginning of the simulation and reaches a steady
state around $t = 10 t_{\rm{p}}$. The total torque decreases in the
adiabatic disc by the saturation, on the other hand, the total torque
remains the same in the optically thick accretion disc. We do not have
to worry about the saturation because we consider the planet in a disc
with the viscosity.

\begin{figure*}
\centerline{\includegraphics[scale=0.6,clip]{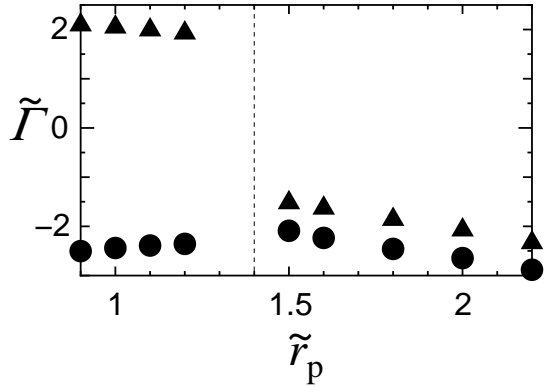}}
\caption{
The total torques exerted on the planet by the disc with
$(\xi_{\rm{gr}}, \xi_{\rm{v}}) = (1.0, 1.0)$ (filled circle) and (10, 0.1)
(filled triangle) as a function of $\tilde{r}_{\rm{p}}$. The boundary
position between the regions~1 and 2 is shown by the dotted
line. In the region~1, the total torque seems to be nearly independent
of $\tilde{r}_{\rm{p}}$. In the region~2, the total torque
decreases a little with increasing $\tilde{r}_{\rm{p}}$.
}
\label{rp-torq}
\end{figure*}

We examine the effect of the boundary of the regions of the disc on
the total torque by changing the position of the planet.
Figure~$\ref{rp-torq}$ shows the total torques acting on the planets
at the different positions in the optically thick accretion discs. The
filled circles and triangles correspond to the total torques on the
planet in the discs with $(\xi_{\rm{gr}}, \xi_{\rm{v}}) = $ (1.0, 1.0) and
(10, 0.1), respectively. The total torque on the planet in the
region~1 is nearly independent of $\tilde{r}_{\rm{p}}$. The effect of
the boundary of the regions on the total torque is very small. The
total torque in the region~2 decreases a little with
$\tilde{r}_{\rm{p}}$. The analytical formula for the normalized
corotation torque \citep{b45} depends on the scale height and the
viscous coefficient, and the corotation torque decreases with
increasing $\tilde{r}_{\rm{p}}$
in the region~2. This agrees with our numerical results.

\begin{figure*}
\centerline{\includegraphics[scale=0.6,clip]{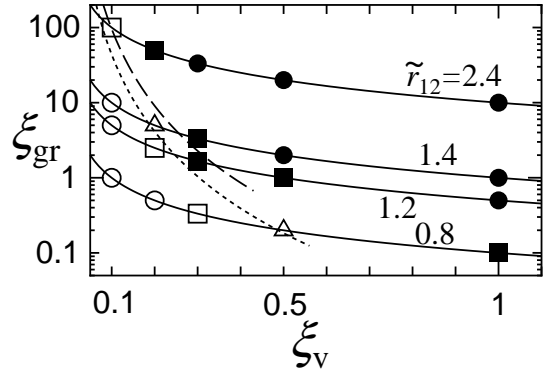}}
\caption{
The diagram of the total torques exerted on the planet with 5~Earth
masses by the disc with various parameter sets of $(\xi_{\rm{gr}},
\xi_{\rm{v}})$. The open and filled marks correspond to the positive
and negative total torques acting on the planets, respectively. We find the
largest magnitude of the total torques, $| \tilde{\Gamma}|_{\rm{max}}
$, from the simulations with the same boundary position of
$\tilde{r}_{12}$. The total torques are plotted as circles, squares,
and triangles when $ | \tilde{\Gamma}| \geq 0.5 | \tilde{\Gamma}
|_{\rm{max}}  $, $0.5 | \tilde{\Gamma} |_{\rm{max}} > |
\tilde{\Gamma}| \geq 0.1 | \tilde{\Gamma} |_{\rm{max}}$, and $0.1 |
\tilde{\Gamma} |_{\rm{max}}  > | \tilde{\Gamma}| $, respectively. The
dotted and dashed curves are drawn using $\tau_{\rm{visc}} =
0.5 \tau_{\rm{turn}}$ when $\tilde{r}_{\rm{p}} = 0.7$ and 1.0,
respectively. It is noted that the curves are truncated because
$\tilde{r}_{12}$ is smaller than $\tilde{r}_{\rm{p}}$. The sign of the
total torque changes when
$\tau_{\rm{visc}} \simeq 0.5 \tau_{\rm{turn}}$. The planet with 5 Earth
masses moves
outward in the region~1 of the optically thick accretion disc with the
surface density of 100~${\rm{g/cm^2}}$ at 1~AU when the
accretion rate is
smaller than $2.1 \times 10^{-8} M_{\odot} / $yr.
The planets are then expected to accumulate at the boundary of the
regions~1 and 2 in the
optically thick disc. }
\label{dmdt-kappa-torq}
\end{figure*}

We perform a number of numerical simulations of gravitational
interactions between the planet and the optically thick accretion
discs. Figure~$\ref{dmdt-kappa-torq}$ shows the magnitude and sign of
the total torques exerted on the planet in the region~1 of the discs
with various parameters of $\xi_{\rm{gr}}$ and $\xi_{\rm{v}}$. The
planet is located at 1~AU when $\tilde{r}_{12} = 2.4$, $1.4$, and 1.2,
and at 0.7~AU when $\tilde{r}_{12} = 0.8$, respectively. The open and filled
marks denote the positive and negative total torques acting on the
planets, respectively. We obtain the largest magnitude of the total
torques, $| \tilde{\Gamma}|_{\rm{max}} $, from the simulations with
the same boundary position of $\tilde{r}_{12}$. The total torques are
plotted as circles, squares, and triangles when $ | \tilde{\Gamma}|
\ge 0.5 | \tilde{\Gamma}|_{\rm{max}}  $, $0.5 |
\tilde{\Gamma}|_{\rm{max}} > | \tilde{\Gamma}| \ge 0.1 |
\tilde{\Gamma}|_{\rm{max}}$, and $0.1 | \tilde{\Gamma}|_{\rm{max}}  >
| \tilde{\Gamma}| $, respectively. The total torque increases with
decreasing $\xi_{\rm{gr}}$ and $\xi_{\rm{v}}$ and its sign changes
from negative to positive.

We find the small decrease in the corotation torque when the viscous
timescale is nearly equal to the turnover time as shown in Fig.~5. The
corotation torque on the planet decreases with the viscosity of the
disc and vanishes when the viscous timescale becomes much shorter than
the turnover time as shown in Fig.~8. We find that the
positive corotation torque cancels the negative Lindblad torque to
have the zero total torque when $\tau_{\rm{visc}} \simeq 0.5
\tau_{\rm{turn}}$. We use the temperature distribution of the region~1
of the disc to find the viscous timescale:
\begin{equation}
\tau_{\rm{visc}} = 1.6 \left( \frac{\xi_{\rm{gr}}}{1.0} \right)^{-1/6}
\left( \frac{\xi_{\rm{v}}}{1.0} \right)^{-7/6} \left(
\frac{M_{\rm{p}}}{5M_{\rm{E}}} \right)
\tilde{r}_{\rm{p}}^{2/3} \Omega_{\rm{p}}^{-1}.
\label{visco_time2}
\end{equation}
Using $\tau_{\rm{visc}} = 0.5 \tau_{\rm{turn}}$, we obtain
\begin{equation}
\xi_{\rm{gr}} =
3.8 \times 10^{-2} \tilde{r}_{\rm{p}}^{11/5} \left(
\frac{M_{\rm{p}}}{5 M_{\rm{E}}} \right)^{18/5} \xi_{\rm{v}}^{-17/5}.
\label{zero-torq-relation}
\end{equation}
The dashed and dot curves in Fig.~$\ref{dmdt-kappa-torq}$ are drawn
using Eq.~($\ref{zero-torq-relation}$). We consider the two positions
of the planet, $\tilde{r}_{\rm{p}} = 1.0$ and $0.7$. Substituting
$\tilde{r}_{\rm{p}} = 1.0$ and $0.7$ into
Eq.~($\ref{zero-torq-relation}$), we draw the dashed and dotted
curves, respectively. The whole region is separated into the two
regions by the curves. The positive total torque and the negative
total torque can be found in one of the regions. We conclude that the
corotation torque is moderately damped by the viscosity when
$\tau_{\rm{visc}} \simeq 0.5 \tau_{\rm{turn}}$ and cancels the negative
Lindblad torque. The opacity of the disc increases with
$\xi_{\rm{gr}}$. Energy dissipates ineffectively in the disc with
large $\xi_{\rm{gr}}$, increasing the mid-plane temperature and the
scale height. The viscous timescale of the disc with large
$\xi_{\rm{gr}}$ decreases because the width of the horseshoe region
decreases with the increasing scale height (see Eq.($\ref{xs}$)). In
the disc with large $\xi_{\rm{gr}}$, the viscosity quickly damps the
corotation torque and the negative Lindblad torque becomes dominant.
Moreover, it is valuable to mention the effect of the surface density 
at 1~AU, $\Sigma_0$, on Eq.~($\ref{zero-torq-relation}$). 
In this study, we set $\Sigma_0$ to be constant. Since the accretion rate and the 
optical depth are proportional to the surface density, $\tau_{\rm{visc}}$ and 
$\tau_{\rm{turn}}$ also increases with $\Sigma_0$. Hence, the coefficient 
of $\xi_{\rm{v}}$ on the right hand side of Eq.~($\ref{zero-torq-relation}$) 
changes by $\Sigma_0$. 

\begin{figure*}
\centerline{\includegraphics[scale=0.6,clip]{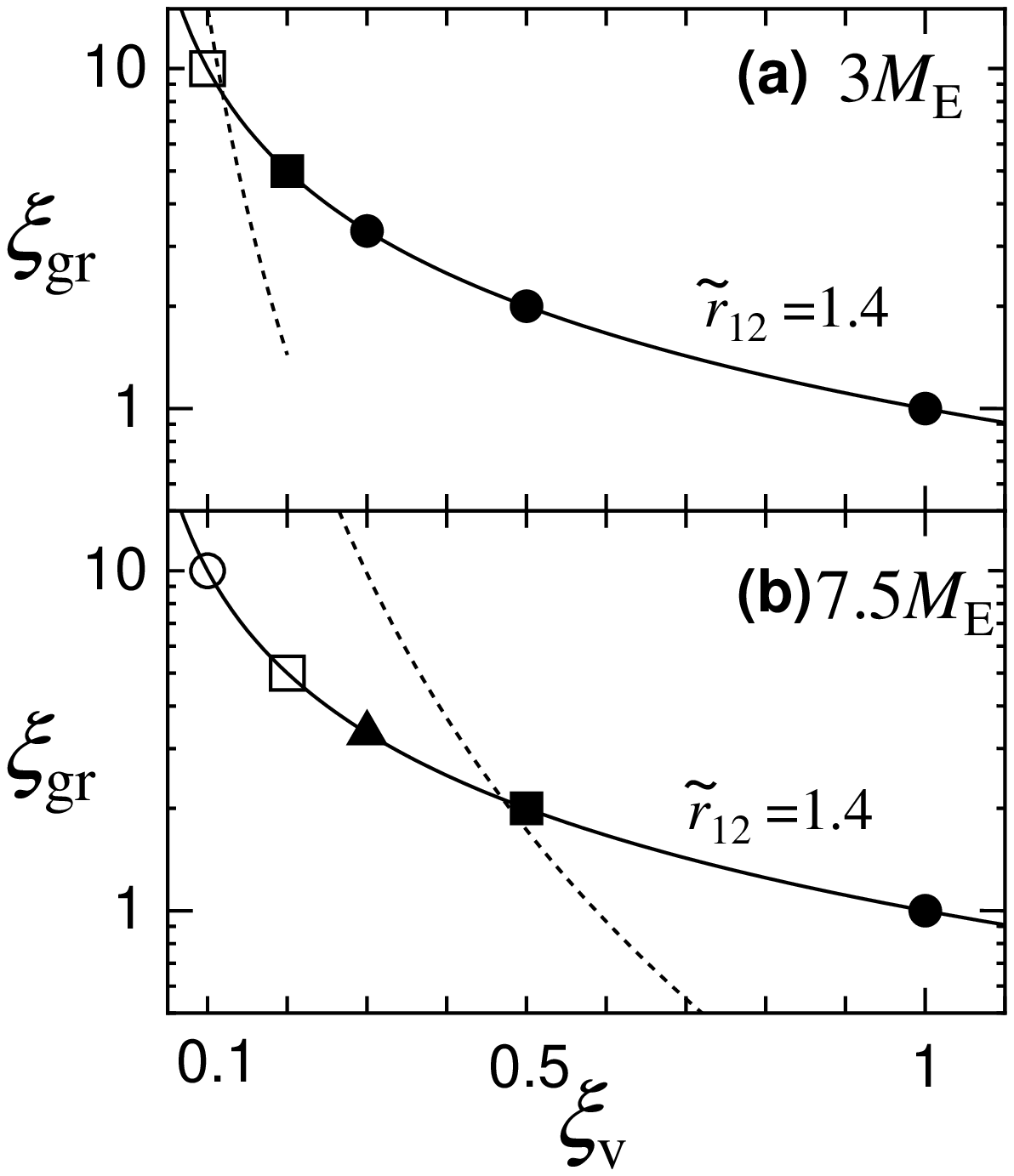}}
\caption{
Same as Fig.~$\ref{dmdt-kappa-torq}$, but the planet with (a) 3~Earth
masses and (b) 7.5~Earth
masses. The open and filled marks correspond to the positive and negative total
torques, respectively. The dotted curve is drawn using
$\tau_{\rm{visc}} = 0.5 \tau_{\rm{turn}}$ when $\tilde{r}_{\rm{p}} =
1.0$. The planets with 3 and 7.5~Earth masses moves outward in the
region~1 of the optically thick accretion disc with the
surface density of 100~${\rm{g/cm^2}}$ at 1~AU when the accretion rate
becomes smaller than $1.2 \times 10^{-8}$ and $3.2 \times 10^{-8}
M_{\odot}/$yr, respectively. }
\label{dmdt-kappa-torq-mp10me}
\end{figure*}

Additionally, one can find from Eq.~($\ref{zero-torq-relation}$) that the planet
mass has a large effect on the threshold at which the total torque
becomes zero. We examine the effect of the planet mass on the total
torque. Figure~$\ref{dmdt-kappa-torq-mp10me}$ shows the total
torque exerted on the planet with 3~Earth masses and that with
7.5~Earth masses in the panels (a) and (b), respectively. The planet is
located at $\tilde{r}_{\rm{p}}=1.0$ in both cases. The half width of the
horseshoe region increases with the planet mass. It takes longer time
to damp the density enhancement in the horseshoe region of the planet
with large mass. The larger viscosity is required to reduce the
corotation torque acting on the massive planet.

\section{Summary}

We have studied the Type~I migration of a planet in an optically thick
accretion disc. The gravitational interactions between a planet and
disc gas excite spiral density waves inside and outside of the orbit
of the planet in the disc. The waves attract the planet
gravitationally and exert torques on the planet. The negative torque
by the outer density wave is a little larger than the positive torque
by the inner density wave because the outer wave is closer to the
planet than the inner wave due to the negative pressure gradient. The
sum of the torques by the density waves (the Lindblad torque) becomes
negative, leading to the inward migration of the planet \citep{b30}.

The corotation torque is very important to planetary
migration because it might be able to halt the inward migration or
reverse its direction \citep{b2, b22}. In a non-barotropic disc, it is
found that the entropy related non-linear corotation torque
plays an essential role. The non-linear corotation torque comes from
the density enhancement due to the generated vortensity at the
outgoing separatrices \citep{b48, b24}. The entropy related corotation
torque is proportional to the radial gradient of the entropy of a
disc. A disc with the steep entropy gradient yields the large positive
corotation torque. \citet{b38} showed that the
positive corotation torque cancels the negative Lindblad torque when
the power-law index of the entropy distribution of the disc becomes
$\lambda = -0.4$, that is the critical power-law index of the entropy
distribution.

The radiation from the central star cannot reach the mid-plane of an
optically thick disc. The temperature of the mid-plane of an optically
thick accretion disc is determined by the energy balance between the
viscous heating and the radiative cooling. The rate of
the cooling by the radiation is sensitive to the opacity of the disc.
Micron size dust particles provide the sources of the opacity. The
opacity changes with the temperature through the
sublimation of ice. We utilize the opacity law frequently used in other
researches and find that the disc is divided into three regions: from
the region~1 with the highest temperature to the region~3 with the
lowest temperature. The boundary positions of the regions are
dependent on the viscosity and the opacity of the disc. The boundary
between the regions~1 and 2 is located between 0.5~AU and 2.5~AU.

Each region has the different power-law indexes of the temperature and
entropy distributions. The power-law indexes of the entropy
distribution in the regions~1 and 3 are lower than the critical
power-law index of $-0.4$, while the power-law index in the region~2 is
higher. We found that
the total torque exerted on the planet by the adiabatic disc becomes
positive in the regions~1 and 3 and becomes negative in the region~2,
as expected. This means that the planet moves outward in the regions~1 and 3,
while it moves inward toward the central star in the region~2. Planets might
accumulate at the boundary between the regions~1 and 2 in the
adiabatic disc.

Dissipative processes change the magnitude of the corotation torque
\citep{b14, b22, b37, b38}. In a viscous disc, the
vortensity is transferred radially as shown in
Fig.~$\ref{vortensitycontour}$. The viscosity is responsible for the
decrease in the corotation torque. The radiation has an impact on the
evolution of the entropy as well and keeps the entropy gradient within
the horseshoe region. The corotation torque depends on these
dissipative processes due to the viscosity and the radiation
\citep{b48}. The positive total torque exerted on the planet
by an adiabatic disc might become negative once some dissipative
processes are included. We include the viscosity and radiation into
the adiabatic disc and calculate the total torque exerted on the
planet in an optically thick accretion disc. We focus on the planet in
the regions~1 and 2. The magnitudes of the opacity and the viscosity
are expressed with the two parameters, $\xi_{\rm{gr}}$ and
$\xi_{\rm{v}}$, respectively. The opacity and the viscosity increase
with $\xi_{\rm{gr}}$ and $\xi_{\rm{v}}$. The large mass
accretion rate of the disc is found in the disc with large
$\xi_{\rm{v}}$. The temperature of the disc increases when the energy
of the disc dissipates ineffectively (large $\xi_{\rm{gr}}$),
resulting in the narrow horseshoe region. The viscosity decreases the
enhancement of the density in the horseshoe region. The corotation
torque gets smaller when the viscosity is larger. The total torque
exerted on the planet by the optically thick accretion disc depends on
the two parameters.

We have made a number of numerical simulations of gravitational
interactions between the planet and the optically thick accretion disc
with various parameter sets of $\xi_{\rm{gr}}$ and $\xi_{\rm{v}}$. The
total torque always becomes negative in the region~2, leading to the
inward migration of the planet. The total torque becomes positive in
the region~1, if the effect of the dissipation is small. The
dissipative processes work effectively and the total torque becomes
zero when the timescale for the viscosity is half of the turnover time
of the planet in the horseshoe orbit, $\tau_{\rm{visc}} =  0.5 \tau_{\rm{turn}}$. This equation is written
with the parameters as $ \xi_{\rm{v}} = 0.4 \tilde{r}_{\rm{p}}^{11/17}
(M_{\rm{p}}/5M_{\rm{E}})^{18/17} (\Sigma_0 / 100{\rm{g/cm^2}})^{7/17}
\xi_{\rm{gr}}^{-5/17}$, considering $\Sigma_0$ as a parameter. In the optically thick accretion disc with
$\xi_{\rm{gr}} \simeq
1.0$ and  the surface density of 100~${\rm{g/cm^2}}$ at 1~AU, the
accretion rate of the disc is required to be smaller than $2.1
\times 10^{-8} M_{\odot}$/yr for the planet to move outward.
Our study suggests that planets with 5~Earth masses might accumulate and drive
further growth of the planets at the boundary between the regions~1
and 2 in the optically thick accretion disc.
The small accretion rate of gas is required for small planets to move outward.

\section*{Acknowledgments}
We are grateful for valuable comments from T.~Tanigawa and the anonymous
referee whose comments help to improve our manuscript. We thank
Nakazawa Nagare Projects members (T.~Tanigawa, A.~Nouda, Y.~Ujiie,
Y.~S.~Yun, and Y.~Yamaguchi) for fruitful comments on the simulation
program. Useful discussion with Y.~Nakagawa is gratefully
acknowledged. This study is supported by the CPS running under the
auspices of the MEXT Global GCOE Program entitled "Foundation of
International Center for Planetary Science." Some of the numerical
simulations were carried out on the general-purpose PC farm at Center
for Computational Astrophysics, CfCA, of National Astronomical
Observatory of Japan.

\appendix

\section{Effect of $\alpha$ on the total torque on a planet}

\begin{figure*}
\centerline{\includegraphics[scale=0.6,clip]{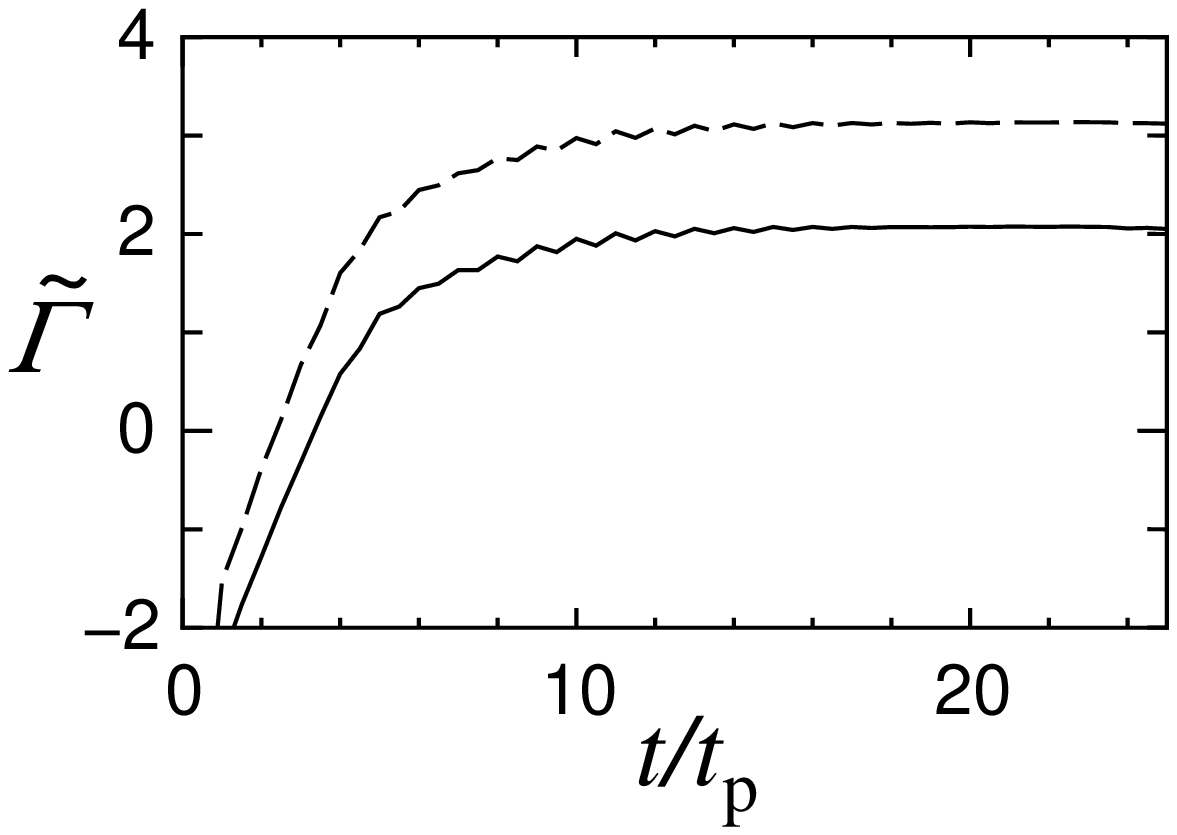}}
\caption{
The time evolution of the total torque exerted on the planet with
5~Earth masses by the disc with $(\xi_{\rm{gr}}, \xi_{\rm{v}}) = (10,
0.1)$. The solid and dashed curves correspond to the total torques
acting on the planet by the discs with $\alpha = 1.0$ and $\alpha =
0.5$, respectively. The planet is located at 1~AU. 
The total torque by the disc with $\alpha = 0.5$
is larger than that by the disc with $\alpha = 1.0$. 
}
\label{alpha-torq}
\end{figure*}

We examine the dependence of the total torque on the power-law index,
$\alpha$, of the surface density distribution of the disc.
Figure~$\ref{alpha-torq}$ shows the time evolutions of the total
torque exerted on the planet by the disc with $\alpha = 1.0$ and the
disc with $\alpha = 0.5$. Both discs have parameters of 
$(\xi_{\rm{gr}}, \xi_{\rm{v}}) = (10, 0.1)$. The solid
and dashed curves represent the total torques acting on the planet by
the discs with $\alpha = 1.0$ and $\alpha = 0.5$, respectively. The
total torques increase at the beginning of the simulations and eventually reach
steady states in both discs. The total torque by the disc with $\alpha
= 0.5$ becomes approximately 1.5~times as large as that by the disc
with $\alpha = 1.0$.

\begin{figure*}
\centerline{\includegraphics[scale=0.6,clip]{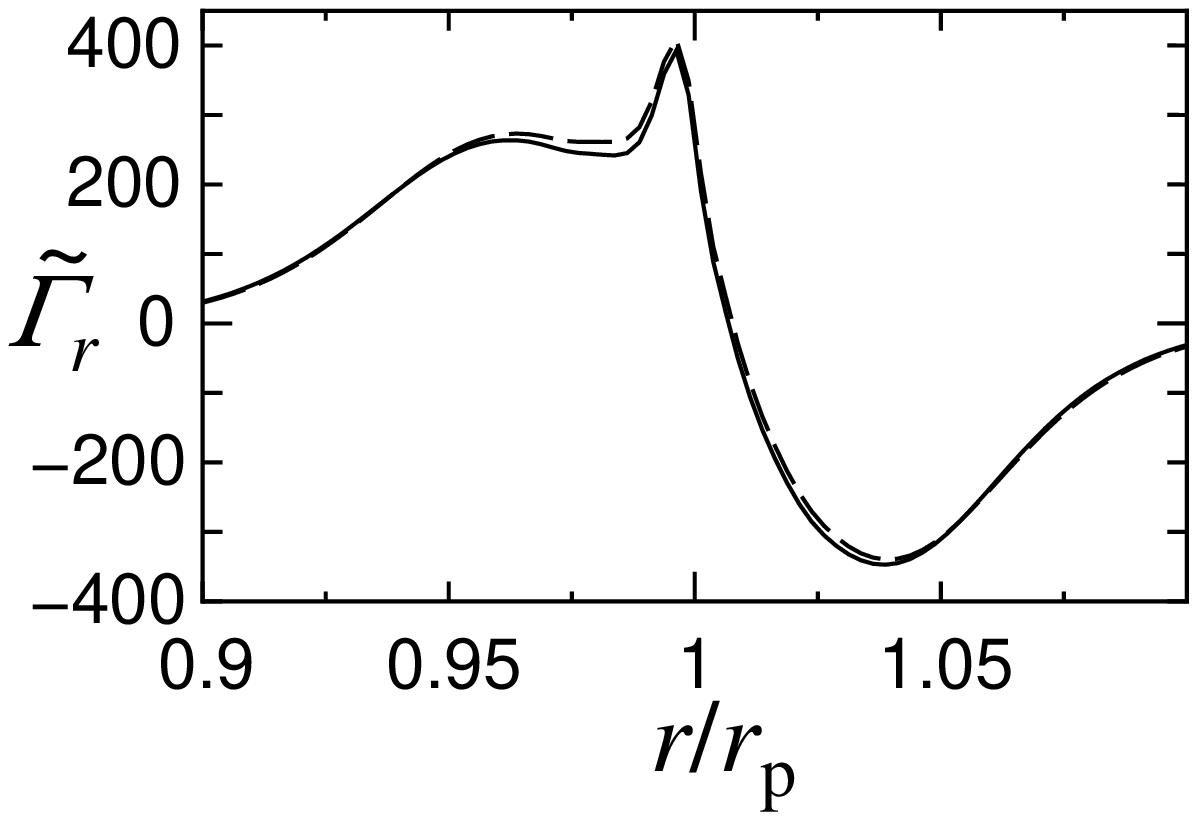}}
\caption{
The torque densities exerted on the planet by the discs with $\alpha =
1.0$ (the solid curve) and $\alpha = 0.5$ (the dashed curve). Both discs
have the parameters of $(\xi_{\rm{gr}}, \xi_{\rm{v}}) = (10, 0.1)$. The planet is located at 1~AU.
}
\label{alpha-torq-den}
\end{figure*}

The torque densities by the disc with $\alpha = 1.0$ and by the disc
with $\alpha = 0.5$ are plotted as the solid and dashed curves,
respectively, in Fig.~$\ref{alpha-torq-den}$. The
inner and outer Lindblad torques increase and decrease with a decrease
in $\alpha$, respectively, leading to the smaller Lindblad torque. The
analytical formula for the Lindblad torque, $\tilde{\Gamma}_{\rm{L}}$, is
given by \citep{b45}
\begin{equation}
\gamma \tilde{\Gamma}_{\rm{L}} = -2.5 - 1.7 \beta_i + 0.1 \alpha.
\label{gammaLno1}
\end{equation}
The temperature distribution of the region~1 of the optically thick
accretion disc has the power-law index, $\beta_1 =
(3+\alpha) / 3$. Hence, we have for the Lindblad torque in the region~1:

\begin{equation}
\gamma \tilde{ \Gamma}_{\rm{L}} = -4.2 - 0.47 \alpha.
\label{gammaLno2}
\end{equation}
The analytical formula suggests that the Lindblad torque
increases with a decrease in $\alpha$ and agrees with numerical
results. The corotation torque becomes a little larger in the disc
with $\alpha = 0.5$, compared with that in the disc with $\alpha = 1.0$.
This increase in the corotation torque is also suggested by the
analytical formula for the corotation torque formula. We can derive the
relation between $\xi_{\rm{gr}}$ and $\xi_{\rm{v}}$ when the total
torque becomes zero, using the same procedure explained in the
section~3.2. Using $\tau_{\rm{visc}} = 0.5 \tau_{\rm{turn}}$, we have
\begin{equation}
\xi_{\rm{gr}} = 3.8 \times 10^{-2} \tilde{r}_{\rm{p}}^{17/10} \left(
\frac{M_{\rm{p}}}{5 M_{\rm{E}}} \right)^{18/5} \left(
\frac{\Sigma_0}{100{\rm{g/cm^2}}} \right)^{7/5} \xi_{\rm{v}}^{-17/5}.
\label{zero-torq-relation-alpha05}
\end{equation}

\newpage

\label{lastpage}


\begin{thebibliography}{99} 
\bibitem[\protect\citeauthoryear{Alibert et al.}{2005}]{b1} Alibert
Y., Mordasini C., Benz, W., Winisdoerffer C., 2005, A\&A, 434, 343
\bibitem[\protect\citeauthoryear{Ayliffe \& Bate}{2010}]{b34} Ayliffe
B. A., Bate M. R., 2010, MNRAS, 408, 876
\bibitem[\protect\citeauthoryear{Ayliffe \& Bate}{2011}]{b33} Ayliffe
B. A., Bate M. R., 2011, MNRAS, 415, 576
\bibitem[\protect\citeauthoryear{Baruteau \& Masset}{2008}]{b2}
Baruteau C., Masset F., 2008, ApJ, 672, 1054
\bibitem[\protect\citeauthoryear{Birnstiel et al.}{2010}]{b35}
Birnstiel T., Dullemond C. P., Brauer F., 2010, A\&A, 513, A79
\bibitem[\protect\citeauthoryear{de Val-Borro et al.}{2006}]{b4} de
Val-Borro M. et al., 2006, MNRAS, 370, 529
\bibitem[\protect\citeauthoryear{Goldreich \& Tremaine}{1979}]{b5}
Goldreich P., Tremaine S., 1979, ApJ, 233, 857
\bibitem[\protect\citeauthoryear{Goldreich \& Tremaine}{1980}]{b6}
Goldreich P., Tremaine S., 1980, ApJ, 241, 425
\bibitem[\protect\citeauthoryear{Hartmann et al.}{1998}]{b7} Hartmann
L.,  Calvet N., Gullbring E., D'Alessio P., 1998, ApJ, 495, 385
\bibitem[\protect\citeauthoryear{Hasegawa \& Pudritz}{2011}]{b46} Hasegawa Y.,
Pudritz R.E., 2011, MNRAS, 417, 1236.
\bibitem[\protect\citeauthoryear{Hubeny}{1990}]{b39} Hubeny I, 1990,
ApJ, 307, 395
\bibitem[\protect\citeauthoryear{Inaba et al.}{2005}]{b12} Inaba S.,
Barge P., Daniel E., Guillard H., 2005, A\&A, 431, 365
\bibitem[\protect\citeauthoryear{Kitamura et al.}{2002}]{b13} Kitamura
Y., Momose M., Yokogawa S., Kawabe R., Tamura M., Ida S., 2002, ApJ, 581, 357
\bibitem[\protect\citeauthoryear{Kley \& Crida}{2008}]{b14} Kley W.,
Crida A., 2008, A\&A, 487, L9
\bibitem[\protect\citeauthoryear{Li et al.}{2000}]{b15} Li H., Finn
J.M., Lovelace R.V.E., Colgate S.A., 2000, ApJ, 533, 1023
\bibitem[\protect\citeauthoryear{Lin \& Papaloizou}{1985}]{b41} Lin D. N. C.,
Papaloizou J., 1985, Protostars and planets II, 981
\bibitem[\protect\citeauthoryear{Lo Curto et al.}{2010}]{b10} Lo
Curto, G. et al., 2010, A\&A, 512, 48L
\bibitem[\protect\citeauthoryear{Masset et al.}{2006}]{b36} Masset F.
S., D'Angelo G., Kley W., 2006, ApJ, 652, 730
\bibitem[\protect\citeauthoryear{Masset \& Casoli}{2009}]{b17} Masset
F., Casoli J., 2009, ApJ, 703, 857
\bibitem[\protect\citeauthoryear{Masset \& Casoli}{2010}]{b48} Masset
F., Casoli J., 2010, ApJ, 723, 1393
\bibitem[\protect\citeauthoryear{Mayor et al.}{2009}]{b16} Mayor M. et
al., 2009, A\&A, 507, 487
\bibitem[\protect\citeauthoryear{Mizuno}{1980}]{b42} Mizuno H., 1980,
Prog. Theor. Phys., 64, 544
\bibitem[\protect\citeauthoryear{Mordasini et al.}{2009}]{b28}
Mordasini C., Alibert Y., Benz W., Naef D., 2009, A\&A, 501, 1161
\bibitem[\protect\citeauthoryear{Morohoshi \& Tanaka}{2003}]{b18}
Morohoshi K., Tanaka H., 2003, MNRAS, 346, 915
\bibitem[\protect\citeauthoryear{Muto \& Inutsuka}{2009}]{b19} Muto
T., Inutsuka S., 2009, ApJ, 701, 18
\bibitem[\protect\citeauthoryear{Paardekooper \& Mellema}{2006}]{b20}
Paardekooper S.-J., Mellema, G., 2006, A\&A, 459, L17
\bibitem[\protect\citeauthoryear{Paardekooper \& Mellema}{2008}]{b21}
Paardekooper S.-J., Mellema, G., 2008, A\&A, 478, 245
\bibitem[\protect\citeauthoryear{Paardekooper \&
Papaloizou}{2008}]{b22} Paardekooper S.-J., Papaloizou J.C.B., 2008,
A\&A, 485, 877
\bibitem[\protect\citeauthoryear{Paardekooper \&
Papaloizou}{2009a}]{b37} Paardekooper S.-J., Papaloizou J.C.B., 2009a,
MNRAS, 394, 2283
\bibitem[\protect\citeauthoryear{Paardekooper \&
Papaloizou}{2009b}]{b23} Paardekooper S.-J., Papaloizou J.C.B., 2009b,
MNRAS, 394, 2297
\bibitem[\protect\citeauthoryear{Paardekooper et al.}{2010}]{b24}
Paardekooper S.-J., Baruteau C., Crida A., Kley W., 2010, MNRAS, 401,
1950
\bibitem[\protect\citeauthoryear{Paardekooper et al.}{2011}]{b45}
Paardekooper S.-J., Baruteau C., Kley W., 2011, MNRAS, 410, 293
\bibitem[\protect\citeauthoryear{Pringle}{1981}]{b43} Pringle J.E., 1981,
Annual Rev. in Astrophys., 19, 137
\bibitem[\protect\citeauthoryear{S\'egransan et al.}{2011}]{b11}
S\'egransan D. et al., 2011, A\&A, 534, 58
\bibitem[\protect\citeauthoryear{Shakura \& Sunyaev}{1973}]{b49}
Shakura, N.I., Sunyaev, R.A.,  1973, A\&A, 24, 337
\bibitem[\protect\citeauthoryear{Tanaka et al.}{2002}]{b25} Tanaka H,
Takeuchi T, Ward W., 2002, ApJ, 565, 1257
\bibitem[\protect\citeauthoryear{Toro}{1999}]{b26} Toro E.F., 1999, in
Riemann solvers and numerical methods for fluid dynamics. A practical
introduction (Springer)
\bibitem[\protect\citeauthoryear{Ward}{1986}]{b27} Ward W., 1986,
Icarus, 67, 164
\bibitem[\protect\citeauthoryear{Ward}{1991}]{b29} Ward W., 1991, LPI, 22, 1463
\bibitem[\protect\citeauthoryear{Ward}{1997}]{b30} Ward W., 1997,
Icarus, 126, 261
\bibitem[\protect\citeauthoryear{Yamada \& Inaba }{2011}]{b38} Yamada
K., Inaba S., 2011, MNRAS, 411, 184

\end{thebibliography}
\end{document}